\documentclass{emulateapj}

\usepackage{color}
\def\etal{~et al.}
\def\simlt{\lower.5ex\hbox{$\; \buildrel < \over \sim \;$}}
\def\simgt{\lower.5ex\hbox{$\; \buildrel > \over \sim \;$}}

\def\gsim{\lower 2pt \hbox{$\, \buildrel {\scriptstyle >}\over
{\scriptstyle \sim}\,$}}
\def\lsim{\lower 2pt \hbox{$\, \buildrel {\scriptstyle <}\over
{\scriptstyle \sim}\,$}}

\def\deg{\ifmmode ^{\circ}
         \else $^{\circ}$\fi}
\def\pdeg{\ifmmode
           $\setbox0=\hbox{$^{\circ}$}\rlap{\hskip.11\wd0 .}$^{\circ}
     \else \setbox0=\hbox{$^{\circ}$}\rlap{\hskip.11\wd0 .}$^{\circ}$\fi}
     
\def\pc{\ifmmode \mathrm{pc} \else $\mathrm{pc}$ \fi}
\def\mpc{\ifmmode \mathrm{Mpc} \else $\mathrm{Mpc}$\fi}
\def\mpcthree{\ifmmode \mathrm{Mpc}^{-3} \else $\mathrm{Mpc}^{-3}$\fi}
\def\gpcthree{\ifmmode \mathrm{Gpc}^{-3} \else $\mathrm{Gpc}^{-3}$\fi}

\def\kelvin{\ifmmode \mathrm{K} \else {$\mathrm{K}$}\fi}
\def\kev{\ifmmode \mathrm{keV} \else $\mathrm{keV}$ \fi}

\def\lsun{\ifmmode {L_\odot} \else $L_\odot$\fi}
\def\msun{\ifmmode M_\odot \else $M_\odot$\fi}
\def\msunyr{\ifmmode M_\odot~\mathrm{yr}^{-1} \else $M_\odot~\mathrm{yr}^{-1}$\fi}

\def\cosi{\ifmmode {\cos\,i} \else $\cos\,i$\fi}

\def\heii{\ifmmode {\rm He{\sc ii}} \else He~{\sc ii}\fi}
\def\mgii{\ifmmode {\rm Mg{\sc ii}} \else Mg~{\sc ii}\fi}
\def\ciii{\ifmmode {\rm C{\sc iii}]} \else C~{\sc iii}]\fi}
\def\civ{\ifmmode {\rm C{\sc iv}} \else C~{\sc iv}\fi}
\def\mgii{\ifmmode {\rm Mg{\sc ii}} \else Mg~{\sc ii}\fi}
\newcommand{\oiii}{{\sc [O~iii]}}

\newcommand{\ovi}{{\sc O~vi}}

\newcommand{\nev}{{[Ne~{\sc v}]}}

\def\teff{\ifmmode {T_{\rm eff}} \else $T_{\rm eff}$\fi}
\def\tmax{\ifmmode {T_{\rm max}} \else $T_{\rm max}$\fi}

\def\mbh{\ifmmode {M_{\rm BH}} \else $M_{\rm BH}$\fi}
\def\led{\ifmmode L_{\mathrm{Ed}} \else $L_{\mathrm{Ed}}$\fi}
\def\lbolflare{\ifmmode L_{\mathrm{bol,flare}} \else $L_{\mathrm{bol,flare}}$\fi}
\def\lagn{\ifmmode L_{\mathrm{agn}} \else $L_{\mathrm{agn}}$\fi}
\def\lbolagn{\ifmmode L_{\mathrm{bol,agn}} \else $L_{\mathrm{bol,agn}}$\fi}
\def\lbol{\ifmmode L_{\mathrm{bol}} \else $L_{\mathrm{bol}}$\fi}
\def\mdot{\ifmmode {\dot M} \else $\dot M$\fi}
\def\mdoto{\ifmmode {\dot{M}_0} \else  $\dot{M}_0$\fi}
\def\mdotf{\ifmmode {\dot{M}_\mathrm{flare}} \else  $\dot{M}_\mathrm{flare}$\fi}

\def\hnot{\ifmmode H_0 \else H$_0$ \fi}

\def\vkep{\ifmmode v_\mathrm{Kep} \else $v_\mathrm{Kep}$ \fi}
\def\vc{\ifmmode v_\mathrm{c} \else $v_\mathrm{c}$ \fi}

\def\vthree{\ifmmode v_{1000} \else $v_{1000}$ \fi}
\def\vrel{\ifmmode v_\mathrm{rel} \else $v_\mathrm{rel}$ \fi}
\def\vkick{\ifmmode v_\mathrm{kick} \else $v_\mathrm{kick}$ \fi}
\def\vkickz{\ifmmode v_{\mathrm{kick},z} \else $v_{\mathrm{kick},z} $ \fi}
\def\vkicky{\ifmmode v_{\mathrm{kick},y} \else $v_{\mathrm{kick},y} $ \fi}
\def\vchar{\ifmmode v_\mathrm{char} \else $v_\mathrm{char}$ \fi}
\def\eflare{\ifmmode E_\mathrm{flare} \else $E_\mathrm{flare}$ \fi}
\def\ekick{\ifmmode E_\mathrm{kick} \else $E_\mathrm{kick}$ \fi}
\def\ecoll{\ifmmode E_\mathrm{coll} \else $E_\mathrm{coll}$ \fi}
\def\ezero{\ifmmode E_\mathrm{0} \else $E_\mathrm{0}$ \fi}
\def\efac{\ifmmode \xi_\mathrm{E} \else $\xi_\mathrm{E}$ \fi}
\def\tqso{\ifmmode t_\mathrm{QSO} \else $t_\mathrm{QSO}$ \fi}
\def\tflare{\ifmmode t_\mathrm{flare} \else $t_\mathrm{flare}$ \fi}
\def\tzero{\ifmmode t_\mathrm{0} \else $t_\mathrm{0}$ \fi}
\def\tfac{\ifmmode \xi_\mathrm{t} \else $\xi_\mathrm{t}$ \fi}
\def\gfac{\ifmmode f_\mathrm{g} \else $f_\mathrm{g}$ \fi}
\def\lflare{\ifmmode L_\mathrm{flare} \else $L_\mathrm{flare}$ \fi}
\def\fflare{\ifmmode F_\mathrm{flare} \else $F_\mathrm{flare}$ \fi}
\def\nflare{\ifmmode N_\mathrm{flare} \else $N_\mathrm{flare}$ \fi}
\def\tshock{\ifmmode T_\mathrm{shock} \else $T_\mathrm{shock}$ \fi}
\def\rmin{\ifmmode R_\mathrm{1} \else $R_\mathrm{1}$ \fi}
\def\rmax{\ifmmode R_\mathrm{2} \else $R_\mathrm{2}$ \fi}
\def\rbound{\ifmmode R_\mathrm{b} \else $R_\mathrm{b}$ \fi}
\def\pbound{\ifmmode P_\mathrm{b} \else $P_\mathrm{b}$ \fi}
\def\mbound{\ifmmode M_\mathrm{b} \else $M_\mathrm{b}$ \fi}
\def\mbo{\ifmmode M_{\mathrm{b}0} \else $M_{\mathrm{b}0} $ \fi}
\def\ebo{\ifmmode E_{\mathrm{b}0} \else $E_{\mathrm{b}0} $ \fi}
\def\efinal{\ifmmode E_\mathrm{final} \else $E_\mathrm{final} $ \fi}
\def\tbound{\ifmmode t_\mathrm{b} \else $t_\mathrm{b}$ \fi}
\def\tagn{\ifmmode t_\mathrm{AGN} \else $t_\mathrm{AGN}$ \fi}
\def\rlim{\ifmmode R_\mathrm{lim} \else $R_\mathrm{lim}$ \fi}
\def\vlim{\ifmmode v_\mathrm{lim} \else $v_\mathrm{lim}$ \fi}
\def\vphi{\ifmmode v_\phi \else $v_\phi$ \fi}
\def\mlim{\ifmmode M_\mathrm{lim} \else $M_\mathrm{lim}$ \fi}
\def\tlim{\ifmmode t_\mathrm{lim} \else $t_\mathrm{lim}$ \fi}
\def\llim{\ifmmode L_\mathrm{lim} \else $L_\mathrm{lim}$ \fi}
\def\fqso{\ifmmode f_\mathrm{QSO} \else $f_\mathrm{QSO}$ \fi}

\def\hbeta{\ifmmode \rm{H}\beta \else H$\beta$\fi}
\def\hbetan{\ifmmode \rm{H}\beta_{\rm n} \else H$\beta_{\rm n}$\fi}

\def\halpha{\ifmmode \rm{H}\alpha \else H$\alpha$\fi}
\def\lalpha{\ifmmode \rm{Ly}\alpha \else Ly$\alpha$}

\def\dvhb{\ifmmode \Delta v_{\hbeta} \else $\Delta v_{\hbeta}$\fi}
\def\dvmg{\ifmmode \Delta v_{\rm{Mg}} \else $\Delta v_{\rm{Mg}}$\fi}

\def\muobs{\ifmmode {\mu_{o}} \else  $\mu_{o}$ \fi}
\def\cosi{\ifmmode {\mathrm{cos}\,i} \else $\mathrm{cos}\,i$\fi}

\def\teff{\ifmmode {T_{eff}} \else $T_{eff}$ \fi}
\def\tmax{\ifmmode {T_{max}} \else $T_{max}$ \fi}

\def\tauh{\ifmmode {\tau_{\rm H}} \else $\tau_{\rm H}$ \fi}

\def\yr{\ifmmode {\rm yr} \else  yr \fi}
\def\kms{\ifmmode \rm km~s^{-1}\else $\rm km~s^{-1}$\fi}
\def\cm{\ifmmode {\rm cm} \else  cm \fi}
\def\cmmitwo{\ifmmode \rm cm^{-2} \else $\rm cm^{-2}$\fi}
\def\cmmithree{\ifmmode \rm cm^{-3} \else $\rm cm^{-3}$\fi}
\def\cmps{\ifmmode \rm cm~s^{-1}\else $\rm cm~s^{-1}$\fi}
\def\cmpsps{\ifmmode \rm cm~s^{-2}\else $\rm cm~s^{-2}$\fi}
\def\kmps{\ifmmode \rm km~s^{-1}\else $\rm km~s^{-1}$\fi}
\def\kmpspmpc{\ifmmode \rm km~s^{-1}~Mpc^{-1} \else
    $\rm km~s^{-1}~Mpc^{-1}$\fi}
  
\def\gcmthree{\ifmmode \rm g~cm^{-3} \else $\rm g~cm^{-3}$\fi}
\def\gcmtwo{\ifmmode \rm g~cm^{-2} \else $\rm g~cm^{-2}$\fi}
   
\def\erg{\ifmmode {\rm erg} \else $\rm erg$ \fi}
\def\ergps{\ifmmode {\rm erg~s^{-1}} \else $\rm erg~s^{-1}$ \fi}
\def\ergcms{\ifmmode \rm erg~cm^{-2}~s^{-1} \else $\rm erg~cm^{-2}~s^{-1}$ \fi}
\def\ergcmshz{\ifmmode \rm erg~s^{-1}~cm^{-2}~Hz^{-1} \else $\rm
erg~cm^{-2}~s^{-1}~Hz^{-1}$ \fi}
\def\ergcmsa{\ifmmode \rm erg~cm^{-2}~s^{-1}~\AA^{-1} \else $\rm
erg~cm^{-2}~s^{-1}~\AA^{-1}$ \fi}
\def\ergshz{\ifmmode \rm erg s^{-1} Hz^{-1} \else
   $\rm erg s^{-1} Hz^{-1}$ \fi}

\def\lam{\ifmmode {\lambda} \else {$\lambda$} \fi}
\def\llam{\ifmmode {L_\lambda} \else  $L_\lambda$ \fi}
\def\lamLlam{\ifmmode \lambda L_{\lambda}(5100) \else {$\lambda L_{\lambda}(5100)$} \fi}
\def\nuLnu{\ifmmode \nu L_{\nu}(5100) \else {$\nu L_{\nu}(5100)$} \fi}
\def\ilam{\ifmmode {I_\lambda} \else  $I_\lambda$ \fi}
\def\flam{\ifmmode {F_\lambda} \else  $F_\lambda$ \fi}
\def\inu{\ifmmode {I_\nu} \else  $I_\nu$ \fi}
\def\fnu{\ifmmode {F_\nu} \else  $F_\nu$ \fi}
\def\bnu{\ifmmode {B_\nu} \else  $B_\nu$ \fi}

\def\msigma{\ifmmode M_{\sigma} \else $M_{\sigma}$\fi}
\def\mbulge{\ifmmode M_{\mathrm{bulge}} \else $M_{\mathrm{bulge}}$\fi}
\def\mgal{\ifmmode M_{\mathrm{gal}} \else $M_{\mathrm{gal}}$\fi}
\def\lgal{\ifmmode L_{\mathrm{gal}} \else $L_{\mathrm{gal}}$\fi}
\def\lbulge{\ifmmode L_{\mathrm{bulge}} \else $L_{\mathrm{bulge}}$\fi}
\def\mgalstar{\ifmmode M^*_{\mathrm{gal}} \else $M^*_{\mathrm{gal}}$\fi}

\def\mbhsigstar{\ifmmode M_{\mathrm{BH}} - \sigma_* \else $M_{\mathrm{BH}} - \sigma_*$ \fi}
\def\deltalogmbh{\ifmmode \Delta~{\mathrm{log}}~M_{\mathrm{BH}} \else $\Delta$~log~$M_{\mathrm{BH}}$\fi}

\def\sigstar{\ifmmode \sigma_* \else $\sigma_*$\fi}
\def\sigthree{\ifmmode \sigma_{\mathrm{[O~III]}} \else $\sigma_{\mathrm{[O~III]}}$\fi}
\def\sigtwo{\ifmmode \sigma_{\mathrm{[O~II]}} \else $\sigma_{\mathrm{[O~II]}}$\fi}
\def\signl{\ifmmode \sigma_{\mathrm{NL}} \else $\sigma_{\mathrm{NL}}$\fi}
\def\wthree{\ifmmode {\rm FWHM({[O~III]})} \else $FWHM({[O~III]})$ \fi}
\def\wtwo{\ifmmode {\rm FWHM({[O~II]})} \else $FWHM({[O~II]})$ \fi}
\def\mthree{\ifmmode M_{\mathrm [O~III]} \else $M_{\mathrm [O~III]}$ \fi}
\def\mtwo{\ifmmode M_{\mathrm [O II]} \else $M_{\mathrm [O II]}$ \fi}

\def\lbreak{\ifmmode L_{\mathrm{break}} \else $L_{\mathrm{break}}$\fi}
\def\lcut{\ifmmode L_{\mathrm{cut}} \else $L_{\mathrm{cut}}$\fi}

\slugcomment{Submitted to ApJ August 1, 2007; revised February 25, 2008.}

\shortauthors{Shields \& Bonning}
\shorttitle{Recoil flares}

\begin{document}

\title{Powerful flares from recoiling black holes in quasars}

\author{G.~A. Shields\altaffilmark{1}, E.~W. Bonning \altaffilmark{1,2}}

\altaffiltext{1}{Department of Astronomy, University of Texas, Austin,
TX 78712; shields@astro.as.utexas.edu} 

\altaffiltext{2}{YCAA - Department of Physics, Yale University, New Haven, CT 06520; erin.bonning@yale.edu}

\begin{abstract}

Mergers of spinning black holes can give recoil velocities from 
gravitational radiation up to several thousand \kms.  A recoiling
supermassive black hole in an AGN retains the inner part of its
accretion disk.  
Marginally bound material rejoining the disk around the moving black hole
releases a large amount of
energy in shocks in a short time, leading to a flare in
thermal soft X-rays with a luminosity approaching the Eddington limit. 
Reprocessing of the X-rays by the infalling material gives strong optical and
ultraviolet emission lines with a distinctive spectrum.   
Despite the short lifetime of the flare ($\sim10^4~\mathrm{yr}$),  as many as $10^2$ 
flares may be in play at the present time in QSOs at redshifts $\sim
1$ to 3.  These flares provide a means to identify high velocity recoils. 

\end{abstract}


\section{Introduction}
\label{sec:intro}

Simulations of binary black hole mergers  show large recoil velocities
(``kicks'') of the final merged black hole resulting from from
anisotropic emission of gravitational radiation [see, e.g.,
\cite{campa07b} for a  summary and references].   
The highest kick so far computed
is $2500~\kmps$ for spin $a_* = 0.8$,
with both black hole spins initially
perpendicular to the orbital angular momentum
and anti-aligned to each other
 \citep{gonzalez07,tichy07}.
\citet{campa07b} predict a maximum recoil velocity of $4000~\kmps$ for
equal mass black holes with maximal spin in this configuration.  For  a binary supermassive
black hole ($\sim10^8~M_\odot$) formed during a galactic merger
\citep{begelman80}, the kick may displace the black hole from the
galactic nucleus or eject it entirely \citep[][and  references
therein]{merritt04}.  For a recoil occurring in an active galactic
nucleus (AGN) with an accretion disk, the inner disk will remain bound to
the black hole, providing fuel for continued AGN activity \citep{loeb07,
bonning07}.   Such a  `wandering QSO' might be observed as a QSO  displaced  from the
galactic nucleus or as a QSO with emission lines shifted relative to
the galactic velocity.  However,  AGN rarely show displaced nuclei
\citep{libeskind06}, and a search for Doppler-shifted broad
 line regions by \citet{bonning07} failed to detect any strong
candidates for recoil.  \citet{bogda07} argue that accretion by the
merging black holes will align their spins in a way unfavorable for
large kicks.  

It is important to explore additional observational signatures of 
high velocity recoils that might help to confirm actual cases
in nature.  We note here that the sudden change in the black hole
velocity leads to an increase in the energy of the disk matter 
that remains bound to the black hole,
evaluated in the reference frame of the moving black hole.  
This can lead to a brief but powerful release of energy in soft X-rays and other forms of
radiation \citep{shields07}. In \S\ref{sec:energetics}, we discuss the mass and energy
involved in the reforming disk and  the time scale and luminosity of the resulting  ``recoil flare.'' 
In \S\ref{sec:nbody} we present a simple numerical simulation of the disk's response to the 
kick.
In \S\ref{sec:appear} we discuss the
observational appearance of the recoil flare, including soft X-ray
emission from the primary shocks and secondary emission from
photoionized gas and heated dust.  In \S\ref{sec:rate} we estimate the
rate of occurrence of these events and the number of recoil flares that may
be observable at the present time.

While this manuscript was undergoing revision, a paper appeared by \citet{lippai08}
also predicting prompt shocks in an accretion disk around
a recoiling black hole.   \citet{lippai08}  focus on smaller radii and
shorter time scales, suitable for searching for optical counterparts
of gravitational wave detections of merging supermassive black holes.
Here we discuss  the outermost bound disk with longer
timescales  with an eye to detecting recoil flares currently in
play in the AGN population.

\section{Energetics}
\label{sec:energetics}

For a black hole merger taking place in an AGN, the accretion disk
fueling the AGN will remain bound to the recoiling  
black hole inside the radius $\rbound = {
(10^{18.12}~\cm) M_8}\,v_{1000}^{-2}$  where the orbital velocity equals
\vkick. Here, $M_8 = \mbh/10^8 M_\odot$ and $v_{1000} =
\vkick/1000$ \kms. For an
$\alpha$ disk \citep{shakura73,frank02}, the disk mass \mbound\ that
remains bound is 
\begin{equation}
\mbound = (10^{6.60} M_\odot) \alpha^{-4/5}_{-1} M^{1/4}_8
 \dot M^{7/10}_{0} R^{5/4}_{17}
\label{eq:mr}
\end{equation}
or
\begin{equation}
\mbound = (10^{8.02} M_\odot) \alpha^{-4/5}_{-1} M^{3/2}_8
 \dot M^{7/10}_{0} v^{-5/2}_{1000} 
\label{eq:mv}
\end{equation}
where $\dot M_{0}$ is the accretion rate in solar masses per
year  \citep{loeb07, bonning07}. 
Here $\dot M$ refers to the normal AGN phase {\em before} the tightening
binary opens a gap in the disk at small radii 
$\sim 10^{2.4} r_g$, where
$r_g \equiv G\mbh/c^2$ \citep{milos05, macfadyen08, loeb07}.
The retained disk mass can fuel QSO activity over a disk consumption time $t_d$
on the order of the viscous timescale of the outer bound disk where most of the mass
resides.   Once the inner gap has refilled, the post-merger accretion rate
will resemble the pre-merger rate, giving
$t_d ~\approx \mbound/\dot M_0 \approx  (10^8~{\rm yr})
\alpha^{-4/5}_{-1} M^{3/2}_8 \dot M^{-3/10}_{0} v^{-5/2}_{1000}$. 

Consider a
recoil directed along the rotation axis (z-axis) of the accretion
disk.  Prior to the recoil, the material at radius  $R$ orbits with
the Keplerian velocity $v_{\rm Kep}$, specific angular momentum $l = R\,
\vkep$, and energy $E = -(1/2)\vkep^2$.   After the kick, the material
still has the same angular momentum around the axis of symmetry;
but it now has a velocity component 
$u_z =  -\vkick$ along the negative z-axis in the reference frame moving with
the black hole.  (We use {\bf u} for velocities in the rest frame of
the post-kick black hole.) This velocity adds in quadrature to the orbital
velocity $u_\phi = \vkep$.  The total energy per unit mass
is now $E^\prime = E + (1/2)\vkick^2$.  This applies at each radius
that remains bound to the hole, and therefore the total energy of the
disk material has increased by an amount 
$\ekick = E^\prime - E = (1/2)\mbound \vkick^2$.   
Because the angular momentum is unchanged, the bound
material will eventually settle into circular orbits at the original radius and binding energy.
The excess energy that the material has immediately after the kick must therefore be dissipated.
We argue below that this dissipation will in large part involve shocks at $\sim \vkick$,
leading to an outburst or ``recoil flare'' with $\eflare = \efac \ekick$, where  \efac\  is of order unity.
The energy associated with a disk with the mass given by Equation \ref{eq:mv} is
\begin{equation}
\eflare =  (10^{57.0}~\erg) \alpha^{-4/5}_{-1}  M^{3/2}_8
  \dot M^{7/10}_{0}  \vthree^{-1/2} \efac.
\label{eq:eflare}
\end{equation}

A more rigorous calculation takes account of the slowing of the black hole as it shares linear
momentum with the bound disk mass.  When the disk has returned to its equilibrium
state, the disk and black hole share a velocity $v_z = \vkick (1 + \mbound/\mbh)^{-1}$;
and the energy dissipated is 
$\ekick  = (1/2)\mbound \vkick^2(1 +  \mbound/\mbh)^{-1}$.  The kinetic energy lost by the black
hole provides the energy radiated and the kinetic energy associated with the motion of
the disk in the $z$-direction.  For $\mbound/\mbh < 1$ as considered here, most of the
energy goes to radiation.    Additional energy will be imparted to disk material outside
\rbound\ that is drawn toward the moving black hole but remains unbound.  We ignore
these complications here. 

Following  a perfectly axial kick, the particles at any radius 
will follow non-intersecting elliptical orbits in the reference frame
of the moving black hole, and a number of orbits  may pass before the
excess energy is dissipated.   
For a kick inclined to the disk axis, the available energy will be similar (see \S \ref{sec:nbody}); but 
particles initially at radii with \vkep\ not greatly larger than \vkick\  (i.e.,
$R$ or order \rbound) will have a post-kick velocity in the black
hole's frame that is a strong function of azimuth in the disk.  These
particles will follow orbits with a range of inclinations and
eccentricities, and the disk will be not merely perturbed but
seriously disrupted.  On approximately the orbital period, material
from different parts of the disk will collide and shock at velocities
of order \vkick.  
Marginally bound material will lag substantially behind the recoiling hole and then
fall back into the disk with velocity $\sim \vkick$.  More tightly bound material in the inner disk
will have its orbits perturbed more modestly and may shock more gently if at all
\citep{loeb07,lippai08}.  However, for the mass surface density underlying Equation \ref{eq:mr},
a majority of the bound mass lies outside $R = 0.5 \rbound$ , and we expect that much of  the excess energy represented by \eflare\ will be dissipated in shocks at velocities of order \vkick 
(see \S \ref{sec:nbody}).
The temperature of an AGN disk at relevant radii is $< 10^4$~K,
with a sound speed $c_s < 10~\kmps$.  Thus,
collisions at   even a fraction of \vkick\ are highly supersonic.

We take the time scale for the recoil flare to be
\begin{equation}
\tflare =  \pbound \tfac =  (10^{3.4}~\yr) M_8 \vthree^{-3} \tfac,
\label{eq:tflare}
\end{equation}
where $\pbound = 2\pi \rbound/\vkick $ is the Keplerian orbital
period at \rbound.   
The factors \efac\ and \tfac\ express the uncertainty
in the radiated  energy and time scale.   Below we estimate 
$\efac \approx 1$ and $\tfac \approx 2$
on the basis of simple numerical simulations.  In \S \ref{sec:appear} we
argue that the flare may be observable for \efac\ as small
as $10^{-2}$.

The power associated with this dissipation is about $\eflare/\tflare$ or
\begin{equation}
\lflare = (10^{46.1}~ \ergps) \alpha_{-1}^{-4/5} M_8^{1/2} \dot M_0^{7/10} \vthree^{5/2} (\efac/\tfac).
\label{eq:lflare}
\end{equation}
For comparison, the bolometric luminosity of the AGN is 
\begin{equation}
\lbol = \epsilon \mdot c^2 \, \gfac = (10^{45.8}~\ergps) \, \mdoto \gfac
\label{eq:lagn}
\end{equation}
for efficiency $\epsilon = 0.1$.  The factor $\gfac \approx 10^{-1}$ allows for suppression of the central accretion rate by the inner gap formed by the binary  \citep{macfadyen08}.  The ratio is
\begin{equation}
\lflare/\lagn = 10^{0.3}\, \tfac^{-1} \gfac^{-1} \alpha_{-1}^{-4/5} M_8^{1/2} \dot M_0^{-3/10} \vthree^{5/2}.
\label{eq:ratiobol}
\end{equation}
Here and below we take $\efac = 1$.
The power of the flare can substantially exceed that of the AGN.  This
is true because a large part of the disk mass shocks on roughly the
orbital period, giving a mass shocked per unit time much larger than the central accretion rate,
which is driven by the comparatively long viscous timescale.
This disparity in time scale offsets the lower energy per unit mass in
the shocks, compared with the black hole accretion efficiency of 
$\sim 0.1 c^2$.

The post-shock temperature will be
\begin{equation}
T_{\rm shock} = (10^{6.9}~\mathrm{K} )(v_{1000})^2 = (0.7~\kev) \vthree^2
\label{eq:tshock}
\end{equation}
\citep{osterbr06}.  The flare lifetime is shorter for higher recoil velocities, but
the luminosity is higher and the X-rays harder.

 The disk mass estimated in Equation \ref{eq:mv} is subject to a number of uncertainties.
 We have extrapolated the standard $\alpha$-disk surface density $\Sigma$ to radii where the disk
 is too cool for the assumed opacity to apply.  However, $\Sigma$ is a weak function of 
opacity \citep{goodman04}.  For larger
\mbh\ and lower \vkick, the stability limit $\mbound < \mbh$ becomes
important  \citep{loeb07}.  If we let \rlim\ and \vlim\ refer to the radius and
orbital velocity within which the $\alpha$-disk mass  equals \mbh,
then from Equation \ref{eq:mr} we have
\begin{equation}
\rlim = (10^{18.11}~\cm) \alpha^{16/25}_{-1} M^{3/5}_8 \dot M^{-14/25}_{0}
\label{eq:rlim}
\end{equation}
and
\begin{equation}
\vlim = (10^{3.01}~\kmps) \alpha^{-8/25}_{-1} M^{1/5}_8 \dot M^{7/25}_{0}.
\label{eq:vlim}
\end{equation}
This limits the energy and luminosity of the flare, but
it remains a powerful event.    

A further complication is the role of the disk's self-gravity
at larger radii \citep{goodman03}. For an $\alpha$-disk, this comes
into play where the Toomre stability parameter reaches unity, $r_Q
\approx 10^{3.6}\, r_g$ and a Keplerian velocity $v_Q \approx
10^{3.7}~\kmps$.   The consequences  for the disk structure outside
$r_Q$ are unclear.   If the surface density of the disk is severely
reduced at radii $\sim 10^5 r_g$ of interest here, then the amount of
shocked gas and the power in the flare will be correspondingly
reduced; but clumping of the disk gas should not greatly affect our
conclusions.   If the disk regulates itself at $Q \approx 1$, 
then Goodman finds a radial dependence of $\Sigma \propto r^{-1}$ or
$r^{-3/2}$ rather than the $r^{-3/4}$ dependence underlying Equation \ref{eq:mr}.  Such
a modification of the disk mass near \rbound\
would affect our quantitative results but leave the recoil flare a readily observable phenomenon.
Goodman summarizes a variety of observational indications that gas disks in AGN do
in fact extend to parsec scales.  We proceed on the assumption that 
a gas disk exists at these radii, but we caution that the mass is uncertain.

\section{Numerical Simulation}
\label{sec:nbody}

In order to test our assumption of high velocity shocks, we have a carried
out a simple numerical simulation of the response of the disk to the
black hole recoil.  We simulated the initial disk as a collection of
collisionless 
particles in Keplerian orbit in the $x - y$ plane, and followed the
motion of the particles and black hole in 3 dimensions with the aid of
an N-body code by \citep{aarseth85}, adapted to our purposes. We
worked in dimensionless units with $G = 1$, a particle mass $m_p
 = 1$,
and $\mbh = 10^5$.   The initial disk had zero thickness.  (An AGN
disk at the radii of interest has $H/R << 1$; and for the non-axial
kicks of interest, post-kick motions out of the original plane far
exceed the initial disk thickness and internal sound speed.)   We
ignored the gravitation force of the disk on itself and on the black
hole.  We modeled the case of a
kick in the $y$--$z$ plane at 45 degrees to the rotation axis.
In the units of our model, the
radial limits of the disk were $\rmin = 10$
and $\rmax = 50$.    We
took $\vkick = 65$, giving $\rbound = 23.6$,  intermediate between \rmin\ and \rmax.
Test calculations reproduced a stable disk around a stationary hole,  and
the nonintersecting nature of orbits for an axial kick.  Note that the
choice of particle mass and radius scale are arbitrary.   The simulation is defined by
the direction of the kick and the ratios $\rmin/\rbound$ and
$\rmax/\rbound$.  Scaling to specific astrophysical parameters is
straightforward using the expressions in \S\ref{sec:energetics}.

In order to test for collisions, we associated a fixed radius $r_p$ of
unity with each particle. The initial disk  comprised concentric
rings of particles, separated by $2 r_p$ in radius and in azimuth.
Thus, the disk particles are initially just touching their neighbors. This
arrangement gave us about 2000 particles. We defined a `collision' to
occur when two particles passed within $r_p$ of each other.

The expected behavior can be visualized in the rest frame of the black
hole just after the kick.  The particles have velocity components
$-\vkicky$ and $-\vkickz$ added to their original orbital velocity.
The total $y$-component in this frame is  $u_y =  -\vkep \sin(\phi) -
\vkicky$, which varies strongly with azimuth $\phi$.   The binding
energy per unit mass in the frame of the hole is now $E = -G\mbh/R +
(1/2)u^2$, where $u$ is the particle's velocity in the rest frame of
the hole.  This  determines the orbital semi-major axis, $a =
2GM/(-E)$,  and period  $P = 2\pi a^{3/2} (GM)^{-1/2}$.  For all radii
in our simulation, the material in a given ring is bound to the hole
for a range of $\phi$ and unbound at other azimuths. 

\begin{figure}[ht]
\begin{center}
\plotone{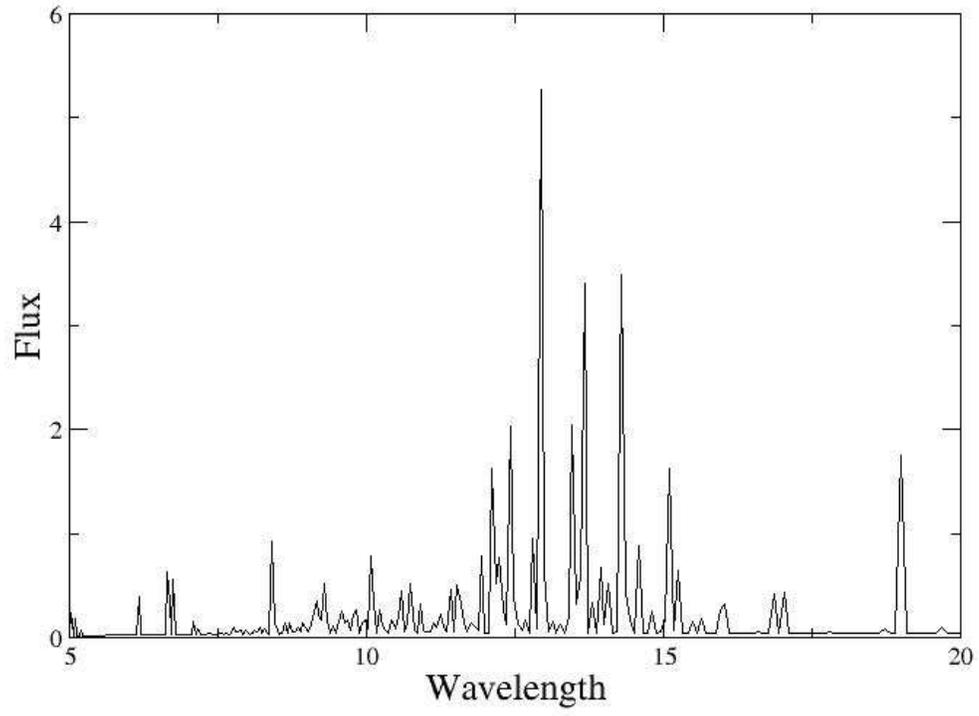}
\figcaption[]{Initial configuration ($t = 0$) of simulated accretion disk before recoil of
the central black hole (see text).  Inner and outer radii are at 10 and 50 in the dimensionless
units of the model.  Disk rotates counterclockwise  seen from above.
\label{fig:i0} }
\end{center}
\end{figure}

\begin{figure}[ht]
\begin{center}
\plotone{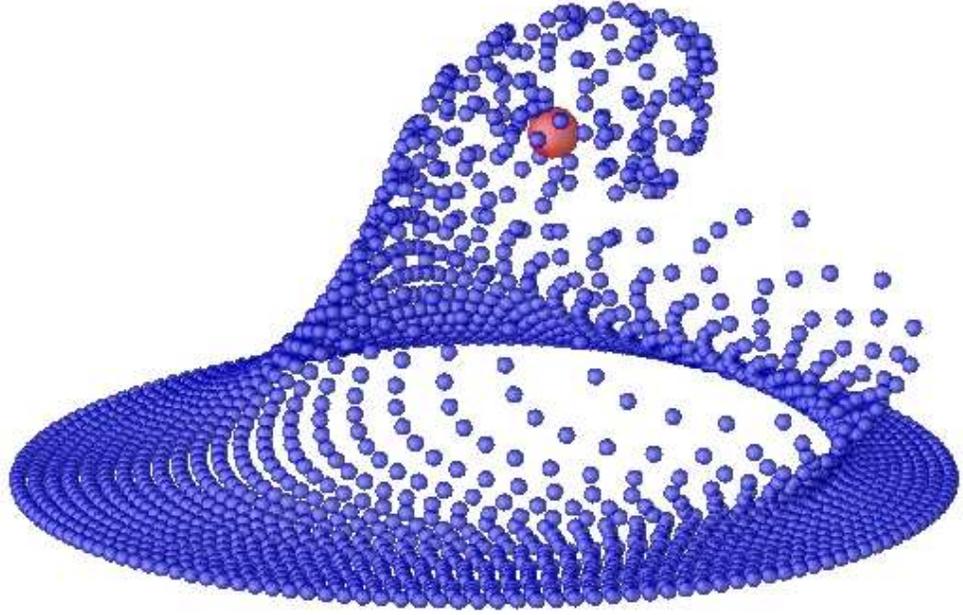}
\figcaption[]{Test particle simulation of accretion disk at time $t = 1$ after recoil of
the central black hole (see text).  In the dimensionless units of the model, the
recoil has $\vkick = 65$ and is inclined 45 degrees to the initial
disk axis. This corresponds to 65\% of the innermost orbital velocity.  The
pre-kick Keplerian orbital period  is $\pbound = 2.28$ at the radius $\rbound = 23.6$ 
where $\vkep = \vkick$. The orbital period at the inner and outer
boundaries of the initial disk are $P = 0.63$ and 7.0, respectively.
\label{fig:i1} }
\end{center}
\end{figure}

\begin{figure}[ht]
\begin{center}
\plotone{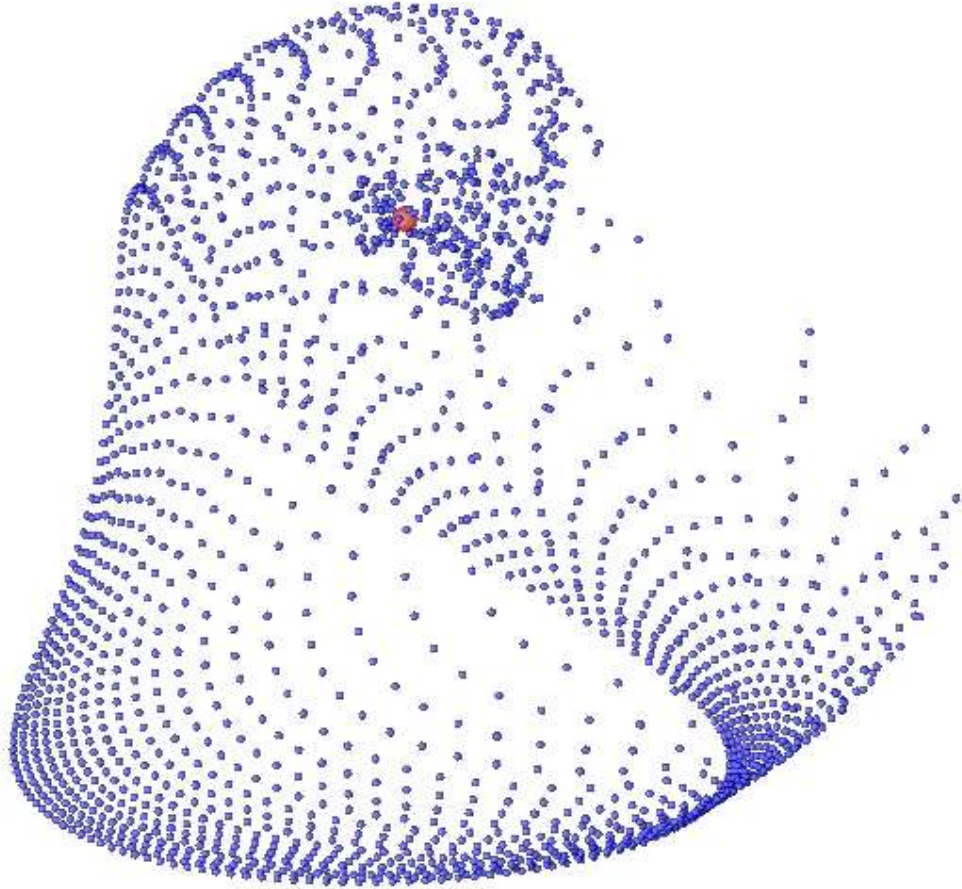}
\figcaption[]{Simulated accretion disk at time $t = 2.5 = 1.1\pbound$ after recoil of
central black hole (see text). 
\label{fig:i2} }
\end{center}
\end{figure}

In the units of our model, the orbital period at \rbound\ is
$\pbound = 2.28$, and this sets the time scale of the recoil flare.
We ran our simulation from $t = 0$ to $t = 14$.
Figures \ref{fig:i0} - \ref{fig:i2}  show the positions of the
particles at several times during the evolution.   
The more tightly bound material at smaller radii follows the hole
relatively closely.  More weakly bound material lags before ultimately
returning to periastron, where it is likely to collide with other
portions of the disk material.  The bound particles experience
collisions at high velocity with other particles originating at
substantially different locations in the original disk.  We take the
energy dissipated per collision to be $(1/2)\vrel^2 m_p$.  

\begin{figure}[ht]
\begin{center}
\plotone{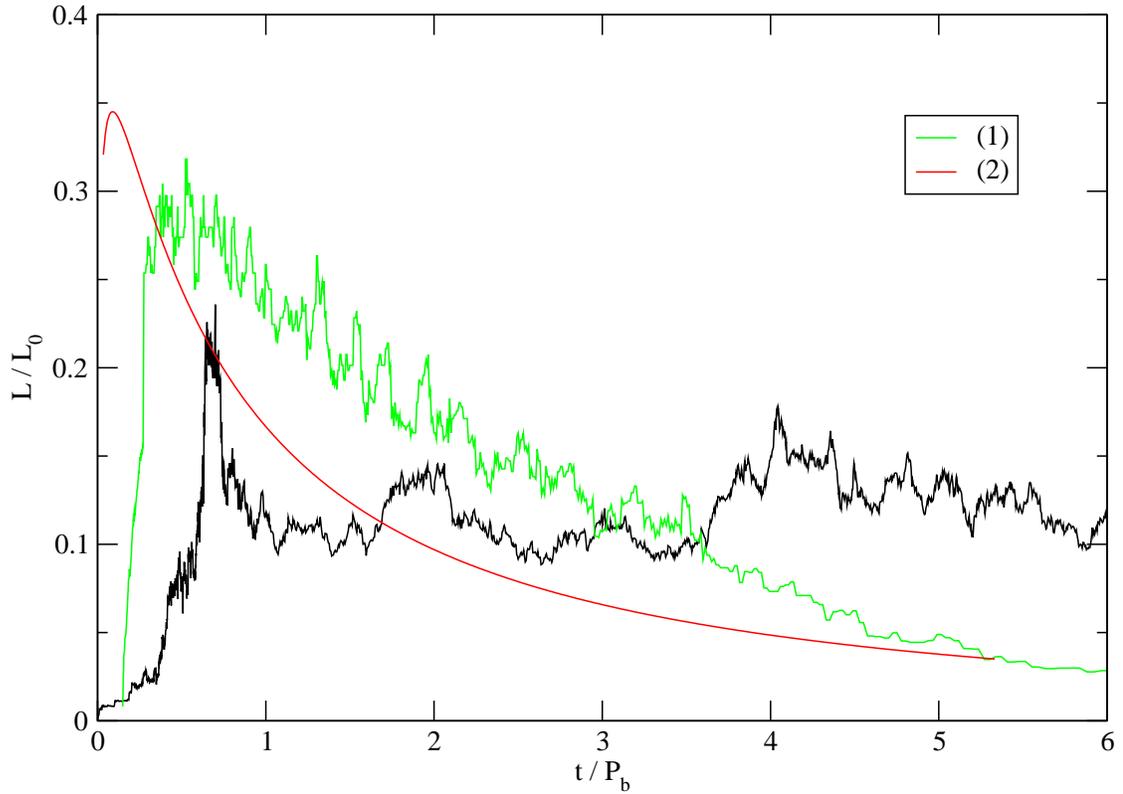}
\figcaption[]{Predicted light curves for recoil flare (see text).
Black curve with sustained luminosity is power per unit time for
particle collisions in numerical simulation; power at later times is overestimated because of
repeat collisions. 
Green curve with declining power is based
on post-kick orbital period $P$ of each bound particle, assuming that an energy $(1/2) m_p  \vkick^2$
is dissipated at $t = P$.  Smooth red curve is analytic solution for axial kick.  Each curve is
normalized to $\pbound$ for the time axis and $L_0 = (1/2) \mbound \vkick^2 \pbound$
for the luminosity axis. 
For reference, $\pbound = 10^{3.4}~\yr$ and
$L_0  = 10^{45.7}~\ergps$  for
$\mbh = 10^8~\msun$, $\mdot = 10^{-0.5}~\msunyr$ and $\vkick = 1000~\kmps$
(Equations \ref{eq:tflare} and \ref{eq:lflare}).
\label{fig:light} }
\end{center}
\end{figure}

Figure \ref{fig:light} shows the power per unit time produced by the colliding particles, 
computed by summing the energy
for 100 consecutive collisions and dividing by the elapsed time.
The power level rises over a time $\sim \pbound$, and persists for
several times \pbound.  The power at later times is exaggerated
because of repeat collisions involving particles that have completed
multiple orbits.   
The r.m.s. relative velocity  of the colliding particles, weighted
by the collision energy, rises from 
$v/\vkick = 0.3$ near the start of the run to 0.9 at $t = \pbound$ and
1.4 for the entire run. This characterizes which shock velocities dominate the
power  and the shock temperature and spectrum.  

Two theoretical light curves are shown for comparison in Figure \ref{fig:light}. 
(1)  Here
each bound particle was assumed to shock exactly one orbital period after the kick, releasing
an energy $(1/2) m_p  \vkick^2$.   The events were ordered in terms of $P$ and binned in
groups of 50 particles to calculate the luminosity.  (2)  The smooth curve in Figure
\ref{fig:light} represents an analytic solution for an axial kick.  The post-kick orbital period
increases monotonically and approaches infinity as $R$ approaches \rbound.  We assume
that the mass  originating  at each radius shocks at velocity \vkick 
at time $t = P$, giving a power
$L = (1/2) \vkick^2 dM/dt$, where $dM/dt = dM/dP = 2\pi \Sigma R dR/dP$, and
$P(R)$ is given by the post-kick binding energy in the frame of the moving black hole.
For constant surface mass density 
$\Sigma$, this gives 
\begin{equation}
\dot{M} = (4/3)(\mbound/\pbound)\zeta^{-5/3}/(\zeta^{-2/3}+1)^3,
\label{eq:dmdt}
\end{equation}
where $\zeta \equiv t/\pbound$.    This decreases as $\dot{M} \propto t^{-5/3}$ at late times,
a qualitatively similar decline to comparison curve (1) above.
This case is given for reference only, as an axial kick will dissipate slowly (see above); and
for smaller radii (having $P << \pbound$) the orbits are less radically
perturbed.   Despite the approximations, the various light curves
agree at the factor-of-two level as to the power expected at times of
one to several times \pbound. 

Integrated over the run, the power dissipated in collisions is $\ecoll
= 2.8\times10^{6}$,  somewhat exaggerated by repeat collisions  at
later times.
We may compare this with the overall energetics of the disk.  Of the
1968 particles in the model disk, 831 remain bound to the hole, so the
bound mass is $\mbo = 831$.  This compares with 381 particles that are
initially inside \rbound; for the tilted kick, a substantial part of
the mass outside \rbound\ remains bound to the black hole.  The
binding energy of the bound mass immediately after the kick is $\ebo =
-1.28\times10^6$, and its total angular momentum  is $L_{b0} =
1.07\times10^6$.  Let us assume conservation of  angular momentum
for the bound mass   in the aggregate as it settles into a circular
disk around the recoiling hole.  
The shallowest binding energy for this mass and
angular momentum occurs for a narrow orbiting ring, giving a radius
$R_\mathrm{ring} = 16.6$ and energy 
$ E_\mathrm{ring} = -2.51\times10^6$.  In this case,  the energy that
must be dissipated to reach this configuration is  $\Delta E =  \ebo
-  E_\mathrm{ring} =  1.23\times10^6$.   Alternatively,  if the final
state is a uniform disk with a sharp outer boundary, then $\efinal =
(32/25)E_\mathrm{ring} = -3.21\times 10^6$  and the dissipated energy
is $\Delta E = 1.93\times10^6$.  These values are comparable to the
simple estimate $\Delta E = (1/2) \mbo \vkick^2 = 1.76\times10^6$,
using here the actual bound mass for the tilted kick.   This supports
a value $\efac \approx 1$ in Equation \ref{eq:eflare}. 

For our model, the most tightly bound quarter of the particles have
a post-kick orbital period $P <  1.1 \pbound$, one half of  all the
bound particles have period $P < 2.4 \pbound$, and 
 the most weakly bound quarter of the particles have $P >  5.4 \pbound$.  This suggests that,
 to characterize the main surge of the flare, we may take $\tfac \approx 2$ in Equation
 \ref{eq:tflare}.  There is some material with arbitrarily weak binding energy after the kick.  This
 material lags far behind the moving hole, giving a prolonged tail to the impact rate.
In our numerical model, the bulk of the shock power corresponds to collisions occurring at radii
near or inside \rbound; the mean collision radius, weighted by collision energy, 
is $\sim12$ or $0.5\rbound$,
with little variation during the run.  
The collisions typically  involve pairs of bound particles 
that originate at substantially different radii near or outside \rbound, and
substantially different azimuths.  These particles move on quite
eccentric orbits, and the collisions  producing most of the power
typically occur not far from periastron. 

We conclude that a non-axial kick will severely disrupt the orbits of
the mass elements in the disk at radii $\sim \rbound$, leading to
strong shocks.  A physically realistic hydrodynamical simulation will be needed to
compute a more detailed light curve.  However, our simulation supports the
approximate energetics outlined in \S \ref{sec:energetics}.

\section{Appearance}
\label{sec:appear}

Equation \ref{eq:tshock} suggests that shocked gas in the post-kick disk
will release its energy largely in the form of thermal soft X-rays,
with $kT \approx 0.7$~keV for $\vkick = 1000~\kmps$, or 0.2~keV for
$\vkick = 500~\kmps$.   Subject to the redshift and  absorption by
the surrounding material, this radiation could be observable.

Consider a QSO with $\mbh = 10^8~\msun$ accreting at
$10^{-0.5}~\msunyr$   (giving $L/\led \approx 10^{-0.8}$) before the
inner gap forms.  Then $\vlim = 750~\kmps$, and for a kick
velocity of $1000~\kmps$ the bound disk mass of $10^{7.7}~\msun$ is
less than \mbh.    The radius of the retained disk in this case
is $\rbound = 10^{18.1}~\cm$, and the surface density is $\Sigma
\approx \mbound/\pi \rbound^2 = 10^{4.2}~\mathrm{ g~cm^{-2}}$.
The above discussion suggests that, for times $t > \pbound$, the recoil flare
has the character of material on highly eccentric orbits impacting the disk
at radii $\sim \rbound$.   
For the parameters just assumed,
 $\tflare \approx \tfac \pbound = (10^{3.4} \yr) \tfac$ or $\sim 5000~\yr$ for
 $\tfac = 2$.  The corresponding mass infall
rate is  $\mdotf \approx \mbound / \tflare = (10^{4.0}~\msunyr)$.
This large rate of mass undergoing shocks is, of course, the basis of
the large power of the flare.
The density of the infalling material is  subject to uncertain geometrical details,
but a simple estimate is
$\rho \approx \mdotf /(2 \pi \rbound ^2 \, v) = 10^{-15.2} \, \gcmthree$, 
or $N = \rho/m_\mathrm{H} = (10^{8.6}~ \cmmithree)$.
The cooling time will be $\sim10^{-2}~\yr$, so that
the material will cool in a relatively narrow zone on the scale of
\rbound.  The local energy flux in the X-ray flare is $\fflare \approx
(1/2)\rho v^3 \approx 10^{8.5}~\ergcms $.
The column density in the impacting stream is
$\sim \rho \rbound \approx 10^{2.9}~\gcmtwo$.  
The optical depth due to photoelectric absorption by H, He, and heavy
elements will be large, and the X-rays may
be unable to penetrate the enveloping inflow.   However, the
luminosity of the shocked flow, $\lflare \approx (10^{45.4}~\ergps)$,
is close to the Eddington limit.  This raises the
possibility of radiation driven outflows, although the
energy available is insufficient to expel the entire disk mass.
Given these outflows and the complex  geometry of the infalling
material, there may be clear lines of sight  on which the X-rays
escape.   While the shock temperature is high, the
effective temperature corresponding to $\fflare$ is only $\teff =
(\fflare/\sigma)^{1/4} \approx  10^{3.2}~\kelvin$.  This underscores
the sensitivity of the observed appearance of the flare to the nature
of the reprocessing that occurs in the surrounding material.

\begin{figure}[ht]
\begin{center}
\plotone{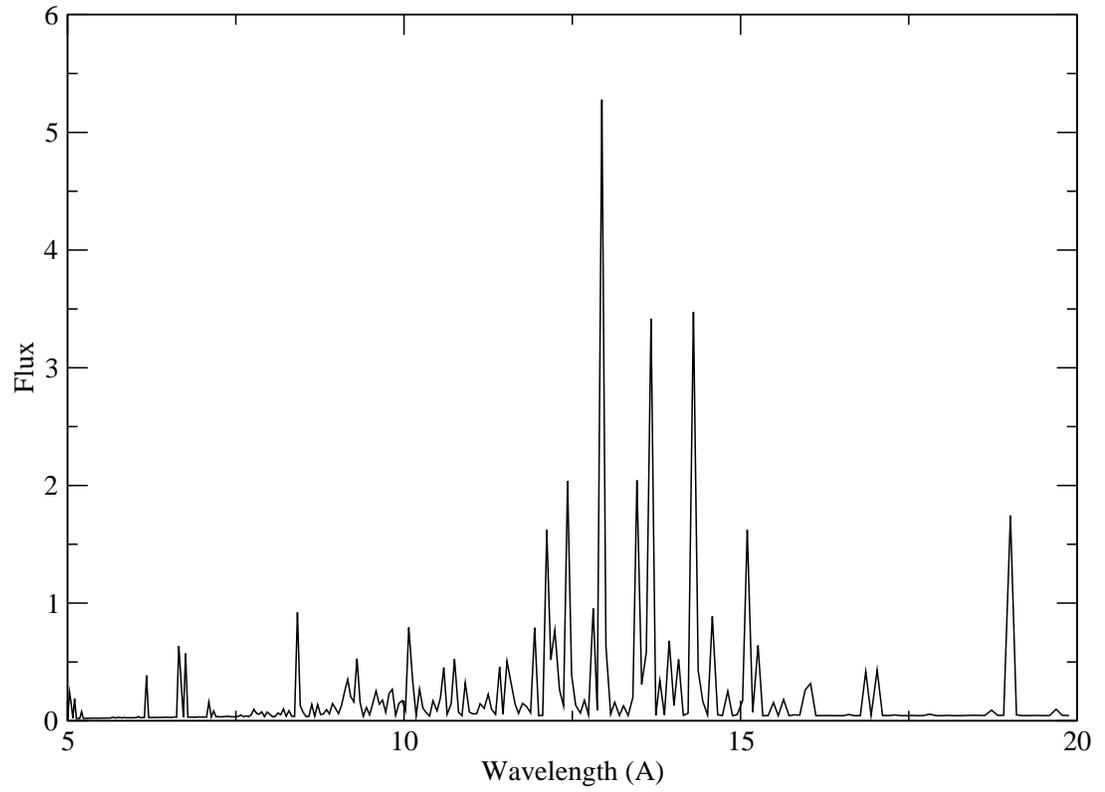}
\figcaption[]{Spectrum of optically thin gas in coronal equilibrium at
$T = 10^7~\kelvin $ in wavelength range 5 to 20 ~\AA\ (see text).
Note strong emission lines of Fe~{\sc xx} and neighboring ions.
Abscissa is wavelength in \AA\ in bins $1500~\kms$ wide; ordinate
is $\lambda F_\lambda$ in arbitrary units.  
The intrinsic line width for the plot is set at 1000~\kms.
\label{fig:corona} }
\end{center}
\end{figure}

For an estimate of the
spectrum of the cooling shocked gas,  we computed a coronal
equilibrium model using  version 07.02.00 of the photoionization code
CLOUDY,  most recently described by \citet{ferland98}.   We used $T =
10^7~\kelvin$, $N = 10^{9.6}~\cmmithree$, and solar abundances.  
(This corresponds to a pre-shock velocity 1100~\kmps\ and
density of $10^{9}~\cmmithree$.) 
Much of the total cooling is thermal bremsstrahlung.  There are many strong
emission lines, including lines of Fe~{\sc XVII} to Fe~{\sc XX} at 12
to 15~\AA, each with several percent of the total energy (see Figure 
\ref{fig:corona}).  If not totally absorbed by the infalling gas, this should
be recognizable as a unique spectrum for a point source in a galactic nucleus.
Comptonization should be insignificant, and the flare will not emit hard X-rays.
There could, however, be hard X-rays if the AGN luminosity is not
entirely shut off by the inner gap in the disk at the time of the merger.

\begin{figure}[ht]
\begin{center}
\plotone{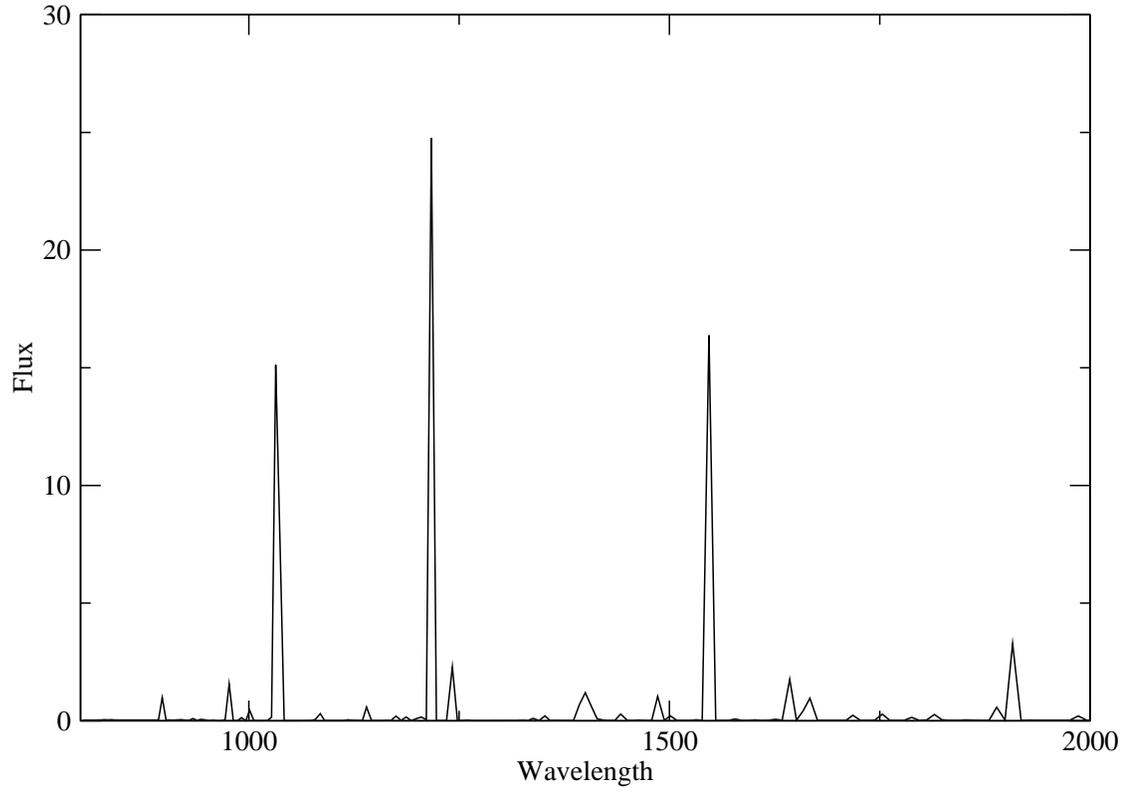}
\figcaption[]{Spectrum of photoionized precursor for
$\tshock = 10^7~\kelvin $ in wavelength range 800 to 2000~\AA\ (see text).
Note strong emission lines of \ovi, \lalpha, and \civ.
Flux is $\lambda F_\lambda$ in arbitrary units.
\label{fig:T7_fuv} }
\end{center}
\end{figure}

\begin{figure}[ht]
\begin{center}
\plotone{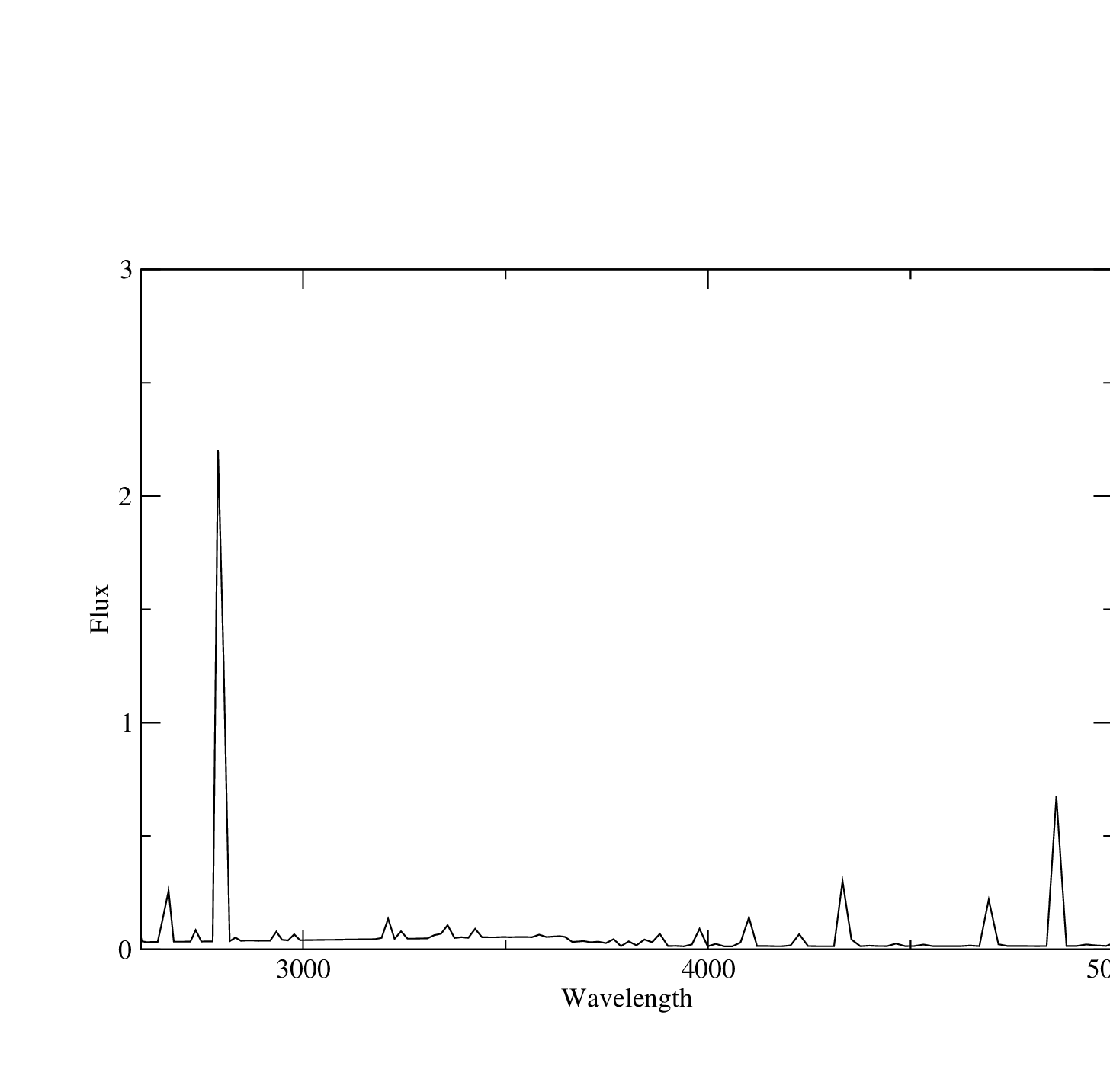}
\figcaption[]{Spectrum of photoionized shock precursor for
$\tshock = 10^7~\kelvin $ in wavelength range 2600 to 5000~\AA\ (see text).
Flux is $\lambda F_\lambda$ in arbitrary units.
Note lines of \hbeta,  \heii~$\lambda4686$, and \mgii,  and 
Balmer continuum in emission.
\label{fig:T7_onuv} }
\end{center}
\end{figure}

The X-ray luminosity will photoionize the infalling material
approaching the shock.  
The local energy flux in the X-ray flare 
directed into the infalling gas is
$\fflare \approx  10^{8.8}~\ergcms$,  giving an ionizing photon flux
of $\phi_i \approx 10^{18.2}~\cmmitwo$ and an ionization parameter
$U = \phi_i/Nc = 0.05$.   We have computed
photoionization models with CLOUDY for this precursor zone.  We used
the coronal model emission as the ionizing spectrum with flux
$10^{8.8}~\ergcms$, a gas density $N = 10^{9}~\cmmithree$, and a
cutoff column density $10^{23.5}~\cmmitwo$.  The photoionized depth  in
this pre-shock flow is $\sim10^{13.7}~\cm$, again narrow compared to
$R_b$.   The high ionization parameter and hard
ionizing spectrum give strong ultraviolet emission lines of highly
ionized species (see Table \ref{t:cloudydeluxe} and Figures \ref{fig:T7_fuv} and
\ref{fig:T7_onuv}).   The \lalpha\ line carries
11\%\ of the incident energy flux, and \ovi\ and \civ\ are comparably strong.
The \hbeta\ and  \civ\  lines respectively have equivalent widths of 
$\sim780$ and 6100~\AA\ in terms of the diffuse
continuum of the precursor alone.  The Balmer and helium lines are affected by large
optical depths.  Based on the coronal equilibrium model, the
shock will contribute an additional 12\%\ to the continuum at \hbeta\ 
and 30\% at $\lambda1550$, slightly reducing the equivalent widths.
The high density suppresses forbidden line emission. 
Aside from this and transfer effects in some lines, 
the spectrum is not strongly sensitive to density, because
the shock velocity alone determines the post shock temperature and also the
ionization parameter of the precursor.

We also computed CLOUDY models for a case corresponding to $\vkick = 500~\kmps$.
We took $M_8 = 10^{-0.5}$ and $\mdoto = 10^{-0.7}$ to conform to the self-gravity limit.
This gives $\tflare = 10^{4.1}~\yr$ for $\tfac = 2$, and a preshock gas density of
$N = 10^{8.1}~\cmmithree$ estimated as above.   Table 1 shows the emission-line intensities of
the precursor ionized by the emission from a $10^{6.3}~\kelvin$ coronal
model for the shocked gas.    Many of the same strong emission lines occur, such as \hbeta, \lalpha,
and \civ.  However, \ovi\ is much weaker, due to a lower level of
ionization and a lower electron temperature.  Thus, the high
ionization line intensities offer a potential diagnostic of the 
shock velocity and hence \vkick.   Forbidden lines such as \oiii~$\lambda5007$, 
\oiii~$\lambda4363$, and \nev\ $\lambda3426$ have intensities of
0.5 to 1.0 times $I(\hbeta)$ because of the lower density.

We may estimate the relative luminosity of the recoil flare at specific wavelengths with the aid
of Equation \ref{eq:ratiobol}.   In the optical, a typical AGN energy distribution gives
$\lamLlam =  0.11 \lbol$ \citep{kaspi00}.   The photoionized precursor by itself has
$\lamLlam =  0.03 \lbol^{flare}$ based on the CLOUDY model.  Using these factors in
equation \ref{eq:ratiobol}, we have
\begin{equation}
\lambda L_\lambda^\mathrm{flare}/\lambda L_\lambda^\mathrm{AGN} = C_\lambda\, \tfac^{-1} \gfac^{-1} \alpha_{-1}^{-4/5} M_8^{1/2} \dot M_0^{-3/10} \vthree^{5/2},
\label{eq:ratioopt}
\end{equation}
with $C_\lambda = 10^{-0.3}$ for $\lambda5100$.  
A similar calculation for $\lambda1550$ gives $C_\lambda = 10^{-0.6}$.
Therefore,  the flare continuum is comparable with
the AGN continuum if $\gfac = 1$ or dominant if \gfac\ is small.   This in turn implies that the equivalent
widths of the flare emission lines  remain large relative to the total observed continuum, 
including the AGN component.   In the composite QSO spectrum
of \citep{vandenberk01}, the
equivalent widths of \hbeta, \civ, \lalpha, and \ovi\  are 46, 24, 93, and 10 \AA, respectively,
mostly  in the broad line component.    
With their large equivalent widths and narrow profiles,
the flare emission lines should be prominent in the total spectrum.  
Indeed,  they remain conspicuous even if the flare
contribution is two orders-of-magnitude weaker than estimated here, relative to the AGN.
The number of detectable flares in play could be increased by 
as much as an order of magnitude by the prolonged arrival of weakly bound material catching up to the black hole (see \S \ref{sec:nbody}).
The line widths will
be $\sim \vkick$, typically 500 to $1000~\kmps$  for flares likely to be observed
(see below).   These line widths resemble those of Narrow Line Seyfert 1 (NLS1) objects
or the  lines from the traditional narrow line region (NLR) of AGN.  
However, the lines of the recoil flare will be distinctive, showing
(1) potentially large equivalent widths, 
(2) weakness or absence of forbidden lines such as \oiii~$\lambda5007$,
(3) strong He~II $\lambda4686$ and possibly O~VI $\lambda1035$, 
and (4) likely velocity shifts and asymmetries of order \vkick.
Absorption by dust in the surrounding disk material may be important.
The disk temperature near \rbound\ is low enough before the kick for dust to exist,  and
the energy flux from the shocks may not be enough to evaporate refractory grains.

We may similarly estimate the prominence of the soft X-rays from the primary shocked gas.
 For the AGN continuum at 0.3~\kev, Laor et al. (1997)
find  an average $\nu L_{\nu,0.3} = 10^{-1.0} \nu L_\nu(3000~\mathrm{\AA})$;
and by the above conversions, this gives 
$\nu L_{\nu,0.3}= 10^{-1.8} \lbol^\mathrm{AGN}$.  From the CLOUDY model for 
$T = 10^7~\kelvin$, we find 
$R_{0.3} \equiv \nu L_{\nu,0.3}^\mathrm{flare}/\lbol^\mathrm{flare} = 10^{-1.0} $.
Combining these expressions with Equation \ref{eq:ratiobol}, we find
%
%\begin{equation}
%\nu L_{\nu,0.3}^\mathrm{flare}  / \nu L_{\nu,0.3}^\mathrm{AGN} 
%= 10^{1.1}\,(R_{0.3}/10^{-1}) \tfac^{-1} \gfac^{-1} 
%\alpha_{-1}^{-4/5} M_8^{1/2} \dot M_0^{-3/10} \vthree^{5/2}.
%\label{eq:ratiox}
%\end{equation}
\begin{eqnarray}
\lefteqn{\nu L_{\nu,0.3}^\mathrm{flare} / \nu L_{\nu,0.3}^\mathrm{AGN}  = } \\
& & 10^{1.1}\,(R_{0.3}/10^{-1}) \tfac^{-1} \gfac^{-1} 
\alpha_{-1}^{-4/5} M_8^{1/2} \dot M_0^{-3/10} \vthree^{5/2}. \nonumber
\label{eq:ratiox}
\end{eqnarray}
If a substantial fraction of the X-ray radiation from the shocks is able to reach the observer, 
the object will stand out as an exceptionally strong source in soft X-rays, relative to its optical and ultraviolet luminosity.   This remains true for
$\vkick = 500~\kmps$, giving $T_\mathrm{shock} = 10^{6.3}~\kelvin$.  Our  coronal equilibrium model at this temperature shows a continuum dropping at $\lambda < 30~\mathrm{\AA}$ 
and many emission lines forming
 a pseudocontinuum at 30 to 60~\AA.   In the vicinity of  40~\AA\ (0.3~\kev), the power in the lines corresponds to $R_{0.3}  = 10^{-0.5} $, again
giving a large ratio of soft X-rays to optical continuum.

The unified model of AGN posits a thick dusty torus that surrounds
the emitting core at a radius of one or several parsecs (Antonucci \&
Miller 1985).  This torus will intercept and reprocess a substantial
fraction of the X-ray and optical-ultraviolet radiation from the
reforming accretion disk.  A large part of the flare luminosity should
therefore appear as infrared radiation from dust.

\section{Rate of Occurrence}
\label{sec:rate}

The number of recoil flares currently observable depends on the rate of
mergers, the probability of kicks of various velocities, and the flare
duration at a given velocity.   Let $dN/dt$ be the merger rate per
year of observer time in some range of \mbh\ and redshift $z$; and let
$f_v$ be the fraction of mergers giving a kick velocity greater than
$v$.  The number of kicks currently observable is roughly $\nflare
\approx (dN/dt) \, [\tflare (1+z)] \, f_v$, where the factor
$1+z$ allows for the observed duration of a flare with intrinsic
duration \tflare.
For mergers of two black holes with equal spin parameter a$_* \equiv
J/M = 0.9$ and roughly equal masses satisfying $q \equiv m_2/m_1 \gsim
0.2$, \citet{schnitt07} find $f_{\rm 500} = 0.31$ and $f_{\rm 1000} = 0.08$. 
These results rely on uncertain
assumptions about the merger details but provide an approximate basis
for estimating the number of high velocity recoils.

Two alternative estimates of $\nflare$ are as follows:

(1)  \citet{sesana04, sesana07}
and \citet{volonteri07} compute binary
merger rates in hierarchical merger simulations normalized to the
observed QSO luminosity function.  Volonteri (2007, personal
communication) has kindly provided merger rates for mass ratios $q  >
0.2$ for  the various seed black hole scenarios of \citet{sesana07}.  
In the ``VHM'' scenario (small seeds), the event rate
per observer year for $1 < z < 3$ is $\sim 0.08~\yr^{-1}$ for  $(m_1 +
m_2)$ in the range $10^7$   to $10^8~\msun$ and $0.014~\yr^{-1}$ in
the range $10^8$ to $10^9~\msun$.  The event rate is about 3 times
smaller in the other scenarios.  The rates drop rapidly for $z < 1$.
Let $\fqso$ be the fraction of these mergers that are sufficiently
gas-rich to fuel QSO activity. 
For  the $10^7$ to $10^8~\msun$ range,
we have $\tflare \approx 800\tfac~\yr$ at $\vkick = 1000~\kmps$ for  $\mbh
=10^{7.5}~\msun$.   Then in the VHM scenario we expect
to see  roughly $ 15\tfac  \fqso$ events currently in play.   For
$\vkick > 500~\kmps$, we expect $\sim 500 \tfac\,\fqso$ events in play,
taking account of the longer flare duration and the larger fraction of
recoils above the lower kick velocity.   For $10^8$ to $10^9~\msun$, the
limit $\mbound < \mbh$ comes into play for a typical value $L/\led = 10^{-0.5}$
(e.g., Salviander et al. 2007), and this affects \tflare.  If we
take $\tflare \approx 2\pi \rlim/\vlim$ (equations \ref{eq:rlim} and \ref{eq:vlim}), 
then we expect  $\sim 8\,\fqso$ flares in play above $1000~\kmps$  and
$\sim 33 \,\fqso$ above $500~\kmps$.  These numbers could be several times
larger if the disk extends beyond \rlim\ with a mass remaining below \mbh.

(2) \citet{haehnelt94} derived the black hole merger rate from the QSO
luminosity function assuming $L=\led$, one event per black hole, and a
QSO lifetime of $t_\mathrm{QSO}= 10^{7.6}~\yr$ (the ``Salpeter time''  for black hole
growth).   At $\mbh$ between $10^7$ and  $10^8~ \msun$,  Haehnelt
finds  $0.016$ events per year of observer time for $z < 3$, giving
roughly 
$3 \tfac$ flares in play  for $\vkick > 1000~\kmps$, and $94 \tfac$  in play
above 500~\kmps .     Uncertainties in this estimate of \nflare\
include the assumed value of $L/\led$ and the QSO lifetime, the flare duration,
and the number of obscured QSOs.  However,  we do not need the factor
\fqso\ because the calculation is based on observed QSOs.

Despite the uncertainties, it  appears likely that several flares
above $10^7~\msun$ are currently observable.  We emphasize that this
rests on the assumption that the black hole merger occurs while the
QSO accretion disk is in place.

For a $10^8~\msun$ hole in a QSO shining at $0.3\led$ at $z = 2$
(prior to formation of the inner gap), the
received flux from the flare is $\sim 10^{-12.8}~\ergcms$ for $\tfac = 2$.  The
observed spectrum will be softened by the redshift, but a considerable
fraction of the radiation should still be received at photon energies
above 0.2 keV.   At $10^{-12.8}~\ergcms$,  \citet{hasinger05}
find  $\sim 2$ soft X-ray sources  (0.5 to 2~\kev) per square degree, 
so that the recoil flares would need to be identified from among
roughly $10^5$  other sources of similar flux.  The unique spectrum of the flare
should aid recognition.

An optical search for recoil flares could target the strong, narrow emission lines
in the rest optical and ultraviolet.    For $\tqso = 10^{7.6}~\yr$, 
the fraction of QSOs in the recoil flare stage from Equation \ref{eq:tflare} is
\begin{equation}
N_\mathrm{flare}/N_\mathrm{QSO} =   10^{-3.2} M_8 \vthree^{-3} \tfac f_v.
\label{eq:duty}
\end{equation}
This gives $N_\mathrm{flare}/N_\mathrm{QSO} = 10^{-4.3} M_8 \tfac$ for
$\vthree = 1$ and $10^{-3.7} M_8 \tfac$ for $\vthree = 0.5$. 
Thus, recoil flares may be observable in about one in 10,000 QSOs, 
and one or more examples may be present in existing large spectroscopic surveys.
For example, the SDSS DR5 QSO Catalog \citep{schneider07} contains
77,429 objects with a median redshift $z = 1.48$ over $\sim 5740~\mathrm{deg}^2$ of sky.  
About $10^4$ of these objects have $z < 0.8$ so that \hbeta\ is accessible, and
many have $z > 1.7$ so that \civ\ is accessible.  The higher redshift objects typically have
relatively high luminosity and black hole masses $\sim 10^{8.5}~\msun$ or higher,
so that the self-gravity limit on the accretion disk is significant.  However, the discussion
in \S \ref{sec:appear} suggests that, even if the disk mass around \rbound\ is
much less than assumed in Equation \ref{eq:lflare}, the flare emission lines
still could be prominent.  
The observed QSO population serves as the parent population for the recoil flares, but
the AGN luminosity may be dimmed several magnitudes by the inner gap at the time of the recoil.
This will complicate the detection of AGN undergoing recoil flares and their inclusion in QSO surveys.

\section{Conclusion}
\label{sec:conclusion}

Merged black holes with high recoil velocities can carry a large mass
of gas in an accretion disk that remains bound to the black hole.  The
kick velocity gives the disk material a boost in energy, relative to
the moving hole, that will in substantial part be dissipated in shocks within
a few orbital periods as the more weakly bound material rejoins the moving disk.
This results in a soft X-ray flare lasting
thousands of years with a luminosity rivaling or exceeding the AGN
luminosity.
The X-rays will be heavily absorbed, at least along some lines of sight,
and the energy re-emitted in line and continuum emission from
photoionized gas and infrared radiation from heated dust.  A number of
recoil flares may be currently observable.  Observational
detection of such an event  would provide an important confirmation of
the occurrence in nature of high velocity recoils.  Detailed
simulations of the observational appearance of these recoil flares are
needed to aid in identifying such objects in surveys at X-ray and
other wavelengths.

\acknowledgments
We thank  Marta Volonteri for providing numerical results and Ski Antonucci, Omer Blaes, 
Richard Matzner, Milos Milosavljevi\'{c}, and Meg Urry for helpful discussions.  EWB 
acknowledges partial support by Marie Curie Incoming European Fellowship contract
MIF1-CT-2005-008762 within the 6th European Community Framework
Programme.

%Table here for mss.

\clearpage

\begin{deluxetable}{lccc}
\tablewidth{0pt}
%\tablewidth{5in}
\tablecaption{Line Intensities of Shock Precursor\label{t:cloudydeluxe}}
\tablehead{
\colhead{Ion}          &
\colhead{$\lambda(\AA)$}       &
\colhead{T7.0}  &
\colhead{T6.3}  \\
}
\startdata

\hbeta	&  4861	&       1.000	&	1.000	\\
\halpha  &  6563	&       5.0		&	6.5		\\
\lalpha  &  1215	&      28.7		&	88.5		\\
He I    &  5876		&       0.10		&	0.24		\\
He II   &  304	         &      17.1		&	19.9		\\
He II   &  1640		&       2.7		&	5.4		\\
He II   &  4686		&       0.31		&	0.68		\\
C II	&  2326		&       1.83		&       0.69		\\
C III	&  1909		&       5.1		&	14.9		\\
C III	&   977		&       2.8		&	2.9		\\
C IV	&  1549		&      25.3		&      55.3           \\
N IV	&  1486		&       1.58		&	5.1		 \\
N V	&  1240		&       3.52		&	1.65		 \\
O III	&  1665		&       2.0		&	4.7		 \\
O IV	&  1402		&       2.1		&	4.8		 \\
O V	&  1218		&       8.6		&	1.6		 \\
O V	&  630		&       5.1		&	0.03		\\
O VI	&  1035		&      48.0		&       0.44		 \\
Mg II	 &  2798		&       5.2		&	5.9		 \\
Si III	&  1397		&       1.6		&	3.6		 \\

\\
log $N_\mathrm{H}$	 &   	 &       9.0	&	8.0	\\   
log $F_\mathrm{inc}$ &   	 &       8.77	&	6.70	\\  
log $F(\hbeta)$  &   	 &       6.34		&	4.21		\\     
EW(\hbeta)    &		 &      780 \AA 	&	660 \AA	\\
EW(\civ)	&  		 &    6100 \AA	&	11,500 \AA		\\
\\

\enddata

\tablecomments{Emission-line intensities relative to \hbeta\ of CLOUDY
models for shock precursor zone photoionized by coronal emission
spectrum at $10^{7.0}$ or $10^{6.3}$~\kelvin (see text).   Incident and
 \hbeta\ flux in \ergcms.  Equivalent widths relative to nebular continuum only.}  

\end{deluxetable}

\clearpage

\clearpage

\clearpage

\clearpage

\clearpage

\clearpage

\clearpage

\end{document}